\documentclass[%
 reprint,
 amsmath,amssymb,
 aps,
]{revtex4-2}
\usepackage{xcolor}
\usepackage{graphicx}% Include figure files
\usepackage{dcolumn}% Align table columns on decimal point
\usepackage{bm}% bold math
\begin{document}

\preprint{APS/123-QED}

\title{Effect of disorder on the strain-tuned charge density wave multicriticality in Pd$_x$ErTe$_3$}% Force line breaks with \\

\author{Anisha G. Singh$^{1,2,3}$, Matthew Krogstad$^{5}$, Maja D. Bachmann$^{1,2,3}$, Paul Thompson$^{6}$, Stephan Rosenkranz$^{5}$, Ray Osborn$^{5}$, Alan Fang$^{1,2,3,4}$, Aharon Kapitulnik$^{1,2,3,4}$, Jong Woo Kim$^{5}$, Philip J. Ryan$^{5}$, Steven A. Kivelson$^{1,2,4}$, and Ian R. Fisher$^{1,2,3}$}
\affiliation{[1] Geballe Laboratory for Advanced Materials, Stanford University, Stanford, CA 94305}
\affiliation{[2] Stanford Institute for Materials and Energy Sciences, SLAC, 2575 Sand Hill Road, Menlo Park, CA 94025}
\affiliation{[3] Department of Applied Physics, Stanford University,  Stanford, CA 94305}
\affiliation{[4] Department of Physics, Stanford University,  Stanford, CA 94305}
\affiliation{[5] Advanced Photon Source, Argonne National Lab, Lemont, IL 60439}
\affiliation{[6] European Synchrotron Radiation Facility, 71 Av. des Martyrs, 38000 Grenoble, France}

\date{\today}% It is always \today, today,

\begin{abstract}
We explore, through a combination of x-ray diffraction and elastoresistivity measurements, the effect of disorder on the strain-tuned charge density wave and associated multicriticality in Pd$_x$ErTe$_3$ (x = 0, 0.01, 0.02 and 0.026). We focus particularly on the behavior near the strain-tuned bicritical point that occurs in pristine ErTe$_3$ (x=0). Our study reveals that while Pd intercalation somewhat broadens the signatures of the CDW phase transitions, the line of first-order transitions at which the CDW reorients as a function of applied strain persists in the presence of disorder and still seemingly terminates at a critical point. The critical point occurs at a lower temperature and a lower strain compared to pristine ErTe$_3$. Similarly, the nematic elastoresistance of Pd$_x$ErTe$_3$, though suppressed in magnitude and broadened relative to that of ErTe$_3$, has a markedly more symmetric response around the critical point. These observations point to disorder driving a reduction in the system's electronic orthorhombicity even while the material remains irrevocably orthorhombic due to the presence of a glide plane in the crystal structure. Disorder, it would appear, reinforces the emergence of a "pseudo-tetragonal" electronic response in this fundamentally orthorhombic material.  
  
\end{abstract}
%\begin{description}
%\end{description}
%\end{abstract}

\maketitle

%\tableofcontents
\section{\label{sec:level1}Introduction}

Erbium tritelluride (ErTe$_3$) is a quasi-2D layered material, hosting a number of distinct strain-tunable unidirectional charge density wave (CDW) states \cite{Ru2006, Straquadine2020, Singh2023}. The goal of the present work is to elucidate the effect of quenched disorder, realized here via Pd-intercalation, on the associated strain-tuned phase-diagram. In particular, we focus on the effect of disorder on the properties proximate to a putative bicritical point associated with the strain-tuned CDW states \cite{Singh2023, Pandey2023}. Our results provide new insights concerning the effect of purposefully introduced disorder to the range of strains and temperatures over which an emergent tetragonal symmetry is apparent near this point.

ErTe$_3$ is a member of the wider family of rare earth tritellurides (RTe$_3$). It is comprised of nearly-square 2D Te bilayers separated by ErTe block layers \cite{Ru_th, Ru2006, Ru2008}. A glide plane between the Te layers renders the material "incurably" orthorhombic (i.e. even if the Te plane is strain-tuned such that the in-plane lattice parameters are equal, the 
crystal is still orthorhombic, lacking a four-fold rotational symmetry). The material undergoes a phase transition at 268 K to a unidirectional, incommensurate CDW state \cite{Ru2006}. In the closely related DyTe$_3$, CDW fluctuations are evident above the critical temperature in both in-plane directions, as can be inferred from the observation of a Kohn anomaly in both directions at the same magnitude wave-vector \cite{Maschek2018}. However, the orthorhombicity of the crystal lattice ensures that the CDW wave-vector always condenses in the same direction in this case along the longer c-axis. (In the standard nomenclature for the space-group \textit{Cmcm}, the a and c-axes lie in-plane, while the b-axis is perpendicular to the Te planes.) 

\begin{figure}
\centering
\includegraphics[scale=0.8]{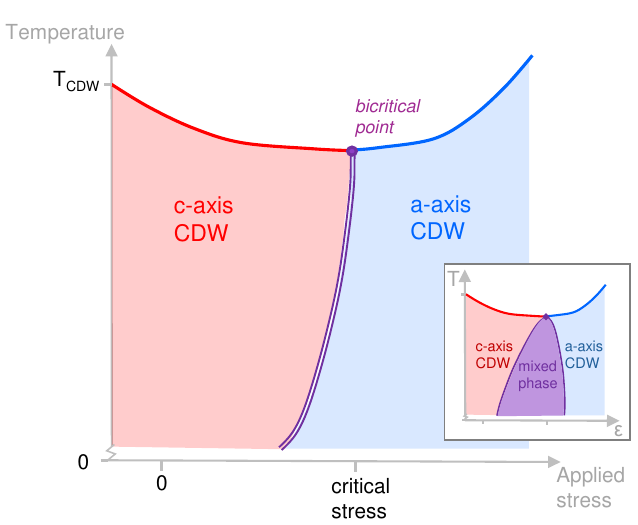}
\caption[Parent Phase Diagram]{\label{fig_1} Schematic diagram illustrating the phase diagram of ErTe$_3$ proximate to the high-temperature CDW phase transition as a function of in-plane uniaxial stress applied along the a-axis. For zero externally applied stress, the material undergoes a phase transition to a unidirectional CDW state with the wave-vector oriented along the c-axis (red phase, labeled "c-axis CDW") at a critical temperature $T_{CDW}$. Beyond a critical value of the stress, however, an a-axis oriented CDW state is observed (blue, labeled "a-axis CDW"). The two phases are separated by a line of first-order phase transitions. Experimental evidence points towards the presence of a bicritical point at the end of this line of first-order phase transitions \cite{Singh2023}. If instead of stress, strain is controlled, the first order transition manifests as a mixed phase region in which domains of both c-axis and a-axis CDW states are observed (purple region in inset) \cite{Singh2023}.} 
\end{figure}

However, the effect of the orthorhombicity on the CDW is weak in that a modest anisotropic strain, induced via external stress, can be used to rotate the direction of the CDW from along the c-axis to along the a-axis (Figure 1)\cite{Straquadine2020, Singh2023}. Since the c and a-axis directions are not related by any symmetry operations of the space group, these two CDW states are fundamentally distinct, even though they are quantitatively similar. The critical stress at which the CDW wavevector suddenly rotates from the c-axis to the a-axis does not correspond to the condition of equal lattice parameters (i.e. does not occur when a = c), and yet, at the bicritical point, where the phase boundaries of the c-axis and a-axis CDW states meet, signatures of an emergent tetragonal symmetry have been detected via elastoresistance measurements \cite{Singh2023, Pandey2023}. In the present work, we explore the effects of chemically-induced quenched disorder on this phase diagram, and on the emergent tetragonality.

Palladium (Pd) intercalation provides a simple means to introduce quenched disorder to ErTe$_3$. Pd impurities behave as isoelectronic intercalants. They do not add any additional charge carriers to the material nor do they substitute into the lattice, but rather they are inserted between the tellurium layers \cite{He2016}. At least for small concentrations, there is no evidence for any ordered superstructure, i.e. the Pd atoms can be assumed to be randomly distributed, creating local deformations that disrupt the perfect periodicity of the lattice \cite{Straquadine2019, Fang2019}. Presumably, these local deformations randomly pin the phase of the CDW, prohibiting the formation of long-range incommensurate CDW order \cite{Larkin1970, ImryMa1975}. Nevertheless, features in transport measurements similar to those observed at the CDW transition in the pristine sample are still observed for Pd$_x$ErTe$_3$, though broader and at lower temperatures and the nature of this low-temperature state and the potential signatures of a phase transition are still open questions \cite{Straquadine2019}. These signatures could 
reflect crossovers that are broadened relics of 
a clean-limit phase transition with no corresponding spontaneous symmetry breaking. A second possibility 
is a phase transition to a Bragg glass phase. Recent x-ray diffraction measurements indicate a vanishing peak width at the transition temperature, suggestive of such a state \cite{Straquadine2019, Mallayya2022}. STM measurements also reveal that dislocations in the CDW state appear in bound pairs, at least for small Pd concentrations, which is also suggestive of a Bragg glass phase \cite{Fang2019}. A third possibility is a phase transition to a state characterized by some kind of vestigial order. For a truly tetragonal material, a vestigial nematic phase would be possible in the presence of weak disorder \cite{Nie2014, Nie2017}. The orthorhombicity of ErTe$_3$ formally precludes such a state in this instance, but the recent observations of broken mirror symmetry in the CDW state \cite{Wang2022a, Alekseev2024, singh2024uncoveringhiddenferroaxialdensity} point to another intriguing possibility. In particular, a vestigial state is possible in which the mirror symmetries spontaneously break even in the absence of long-range incommensurate density wave order (i.e. a vestigial monoclinic or ferroaxial phase). We also note that from a theoretical basis, the issue of whether a bicritical point is even allowed beyond mean-field theory is technically still an open question \cite{Aharony2, Arguello2014}. 

The above questions motivate an experimental study of how the phase diagram for pristine ErTe$_3$ (Figure 1) evolves with Pd intercalation. Our study reveals that while the CDW phase transitions appear to be somewhat broadened, the line of first-order transitions, at which the CDW reorients, persists in the presence of disorder and still seemingly terminates at a critical point.  However, it is unclear whether this is still a bicritical point (appropriate in the case of vestigial order) or a critical endpoint (as would be expected in the case of disorder). Moreover, proximate to this critical point, there remain signatures of an emergent tetragonal symmetry as witnessed by the elastoresistivity. Furthermore, we find that the anisotropy of the disordered material is considerably reduced from that of the pristine compound, such that the critical stress associated with tuning to the critical point is smaller than for the parent compound. Similarly, the regime over which pseudo-tetragonal behavior is observed near the critical point is enlarged relative to the pristine compound, and a greater symmetry between the a-axis and c-axis resistivity and elastoresistivity is also observed.

\section{\label{sec:level2}Methods and Materials}

\subsection{\label{sec:level2.1}Crystal Growth and Compositional Analysis}
Single crystals of ErTe$_3$ and Pd$_x$ErTe$_3$ were grown via standard self-flux methods \cite{Ru2006, Ru_th}. Previous microprobe studies of Pd$_x$ErTe$_3$ crystals \cite{Straquadine2019} revealed that the resulting palladium concentration, \textit{x}, tracks closely with the starting Pd:Te ratio in the growth melt. This relationship was used as a guide to determine the amount of palladium that should be added to a growth to produce a given intercalation level. Though this relationship is quite accurate, especially below x=2$\%$, there can still be a subtle discrepancy between the nominal palladium concentration and the actual palladium concentration in a given sample batch. To establish the palladium intercalation for a given batch, the CDW crossover temperature was measured with transport since the relationship between Pd intercalation and this temperature has been previously established \cite{Straquadine2019}. We explicitly study compositions with x = 0.010, 0.020 and 0.026. 

\subsection{\label{sec:level2.2} Strain Measurement}
\begin{figure}
\includegraphics[scale=0.8]{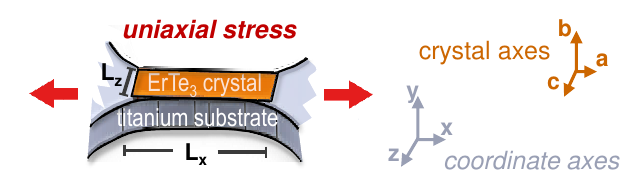}
\caption[Diagram of Strain Measurement]{\label{fig_2} Schematic diagram illustrating the experimental geometry used for strain measurements, and definition of the dimensions $L_x$ and $L_z$ used in the data analysis. The single crystal of Pd$_x$ErTe$_3$ (gold rectangle) is bonded to the thin bridge of a Ti bowtie (grey), with the a-axis orientated along the length of the bar. Uniaxial stress is applied to the Ti bowtie, resulting in anisotropic in-plane strain. Coordinate axes used to describe the measurements are also labeled.} 
\end{figure}

A CS-130 Razorbill strain cell was utilized to apply variable strains to the samples \cite{Hicks2014}. Since RTe$_3$ are micaceous crystals, to maintain their crystallinity during strain cycling, the sample was mounted onto a titanium platform designed for strain transmission rather than directly mounted into the strain cell. This experimental set-up is illustrated schematically in Figure \ref{fig_2}. The \textit{a-c} crystal axes correspond to the \textit{x-z} experimental axes respectively. Samples were cleaved to thicknesses less than 20$\mu m$ to avoid strain relaxation across the height of the sample.  Angstrom Bond epoxy was used for sample bonding to produce a thin, stiff glue layer.

For the strain measurements, the sample was oriented such that the a-axis was along the direction of applied stress, such that tensile strain would result in rotation of the CDW wavevector from along the c-axis to along the a-axis \cite{Singh2023, Gallo2023}. X-ray diffraction was used to align the crystals prior to mounting. During this alignment procedure, samples were also screened to verify they did not have any stacking faults \cite{Josh_th, Singh_th}.

During experiments, the strain state of the sample was determined in two different ways. For x-ray measurements, the value of the lattice parameters can be determined directly from diffraction, and the strain deduced accordingly from the orthorhombicity  \small $\dfrac{a-c}{a+c}$\normalsize. With reference to Figure 1, for strains that result in a monodomain state, this ratio is readily determined. However, for compositions in the mixed phase regime (purple region in the inset to Figure 1), the average strain of the material (distinct from the strain in each domain) must be determined from the weighted average of the lattice parameters. Specifically, rather than using the position of the Bragg peak to calculate the lattice parameter in each domain, which does not change with strain in the mixed phase regime  \cite{Singh2023}, the peak's center of mass was used, which gives an average value of the lattice parameters along the \textit{x}, $<a>$,  or \textit{z}, $<c>$, directions. From these values, the ratio \small $\dfrac{<a> -<c>}{<a>+<c>}$\normalsize can be determined, corresponding to the average strain experienced by the macroscopic sample. 

In contrast, for transport measurements, the strains experienced by the platform $\Delta L_x/ L_x$ and $\Delta L_z/ L_z$ were estimated. $L_x$ is the effective length of the platform, determined for the specific bowtie platforms used in these experiments to be approximately 3.47mm based on numerical simulations \cite{Singh2023, Park2020}. A capacitive sensor in the strain cell measures the displacement between the cell jaws. For small strains, specifically strains lower than the plastic deformation limit of titanium, it is a reasonable assumption that this displacement is completely transmitted to the platform (at least, up to the strain loss associated with the epoxy that bonds the Ti substrate to the cell \cite{Hicks2014}), giving an indirect estimate of $\Delta L_x$. For larger strains, it is more accurate to measure $\Delta L_x$ directly by mounting a strain gauge on to the platform itself. The platform strain, $\Delta L_x/ L_x$ is a reasonable approximation for the macroscopic strain experienced by the sample since the sample geometry is designed for high strain transmission \cite{Singh2023}. The transverse strain of the platform $\Delta L_z/L_z$, determined via Poisson's ratio for Ti, sets the transverse strain of the sample.

\subsection{\label{sec:level2.3}XRD Measurements}

The large reciprocal space maps presented in Section \ref{sec:level3.1} were collected at Sector 6-ID-D at the Advanced Photon Source using methods outlined in the supplement of \cite{Krogstad:2019tc}. At Sector 6-ID-D, a new methodology has been developed to efficiently collect total x-ray scattering over large volumes of reciprocal space. In each measurement, a crystal is rotated continuously through a full revolution at a rate of 1 degree/s while images are collected on a fast area detector every 0.1 s, with an incident monochromatic x-ray beam at 87 keV. This is repeated three times at different detector and goniometer settings to improve statistics and redundancy. These images are then transformed into reciprocal space meshes which include thousands of Brillouin Zones.

The x-ray diffraction (XRD) measurements under strain in Section IIIB were conducted at the Advanced Photon Source in Sector 6-ID-B.  Here, a specialized sample environment has been developed to allow high-resolution XRD measurement with in-situ strain tuning at cryogenic temperatures \cite{Sanchez2021, Malinowski2020}. These measurements were taken at 11.2 keV, the penetration depth of which in ErTe$_3$ is far greater than the thickness of the samples studied. Additional measurements were taken at the European Synchrotron Radiation Facility at the XMaS BM28 beamline.

\subsection{\label{sec:level2.4} Transport Measurements}

To perform transport measurements for samples mounted on a bowtie substrate, it was first necessary to create on insulating oxide layer on the Ti such that the sample resistivity was not shorted by the metallic platform. To do this, the Ti platforms were heated to 750 C in air for 2 hours. Samples were then bonded to the anodized surface using Angstrom bond as described above. Transport experiments in this work focus on measuring the in-plane resistivity anisotropy, $(\rho_a -\rho_c)$, because this quantity is very sensitive to the orientation of the CDW order \cite{Straquadine2020}. A transverse sample geometry was utilized to allow direct measurement of the resistive anisotropy in a single measurement \cite{Walmsley2020}.  The geometry was defined using a Focused Ion Beam (FIB). Further details on the development of the microstructure design can be found in \cite{Singh_th, Singh2023}.

\section{\label{sec:level3}Experimental Results}
\subsection{\label{sec:level3.1}XRD Measurements of Unstrained Samples}

To evaluate the effects of strain on a Pd$_x$ErTe$_3$ sample, it is first necessary to establish the structure of unstrained Pd$_x$ErTe$_3$. Large-scale reciprocal maps can be used to evaluate what change, if any, intercalation elicits in the material's crystal symmetry. We start by examining the crystal structure above the critical temperature, and then examine the lattice modulations in the CDW state. 

\begin{figure}
\centering
\includegraphics[scale=0.588]{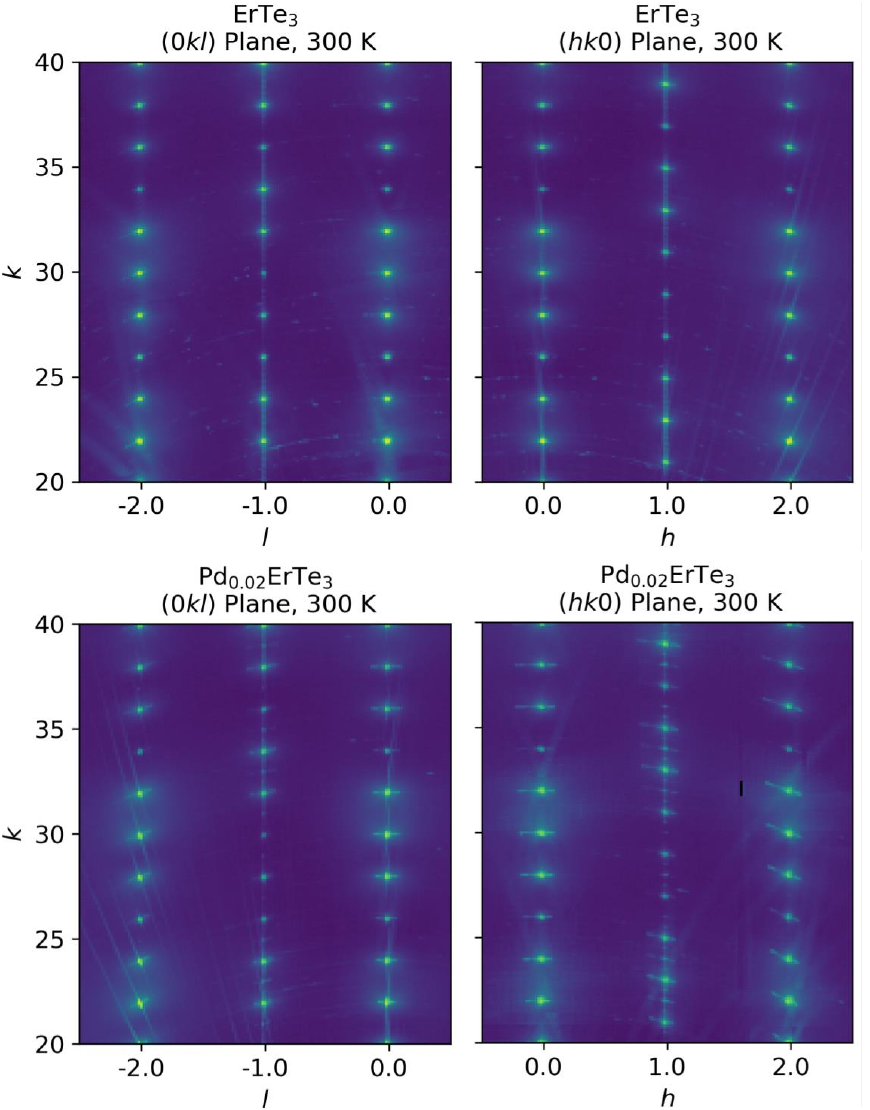}
\caption[Crystal Symmetry of Pd$_x$ErTe$_3$]{\label{fig_4_0} Reciprocal space maps of pristine ErTe$_3$ (top row) and Pd-intercalated Pd$_{0.02}$ErTe$_3$ (bottom row) at 300K, demonstrating features of the crystal symmetry of the material above $T_{CDW}$. Left hand column shows the (0kl) plane, while the right hand column shows the (hk0) plane. Both images in a row share the same y-axis scale, indicated on the left hand side plot. These and similar related data reveal that Pd intercalation does not change the average crystal structure of ErTe$_3$. } 
\end{figure}

The \textit{Cmcm} orthorhombic space group, to which all the rare earth tritellurides belong for temperatures above $T_{CDW}$, is characterized by systematic absences in diffraction measurements. In Figure \ref{fig_4_0}, reciprocal space maps in the $(0, k, l)$ and $(h, k, 0)$ planes are presented for a sample with no intercalation (top row) and one with $x= 2\%$ palladium intercalation (bottom row), in both cases at a temperature of 300 K (i.e. well above $T_{CDW}$). For the \textit{Cmcm} space group, if $h=0$ then peaks are allowed at even values of $k$ regardless of the value of $l$. For example, looking at the $(0,k,l)$ scattering plane of ErTe$_3$, regardless of the value of $l$, the values of $k$ for which a peak is observed are the same. In contrast, for the $(h,k,0)$ plane, the values of $k$ for which a peak is observed are dependent on the value of $h$. For both ErTe$_3$ and Pd$_{0.02}$ErTe$_3$, the average structure (i.e. neglecting the diffuse scattering described below) can be described by the orthorhombic \textit{Cmcm} space group. In other words, Pd intercalation does not change the average crystal structure.  

\begin{figure}
\centering
\includegraphics[scale=0.7]{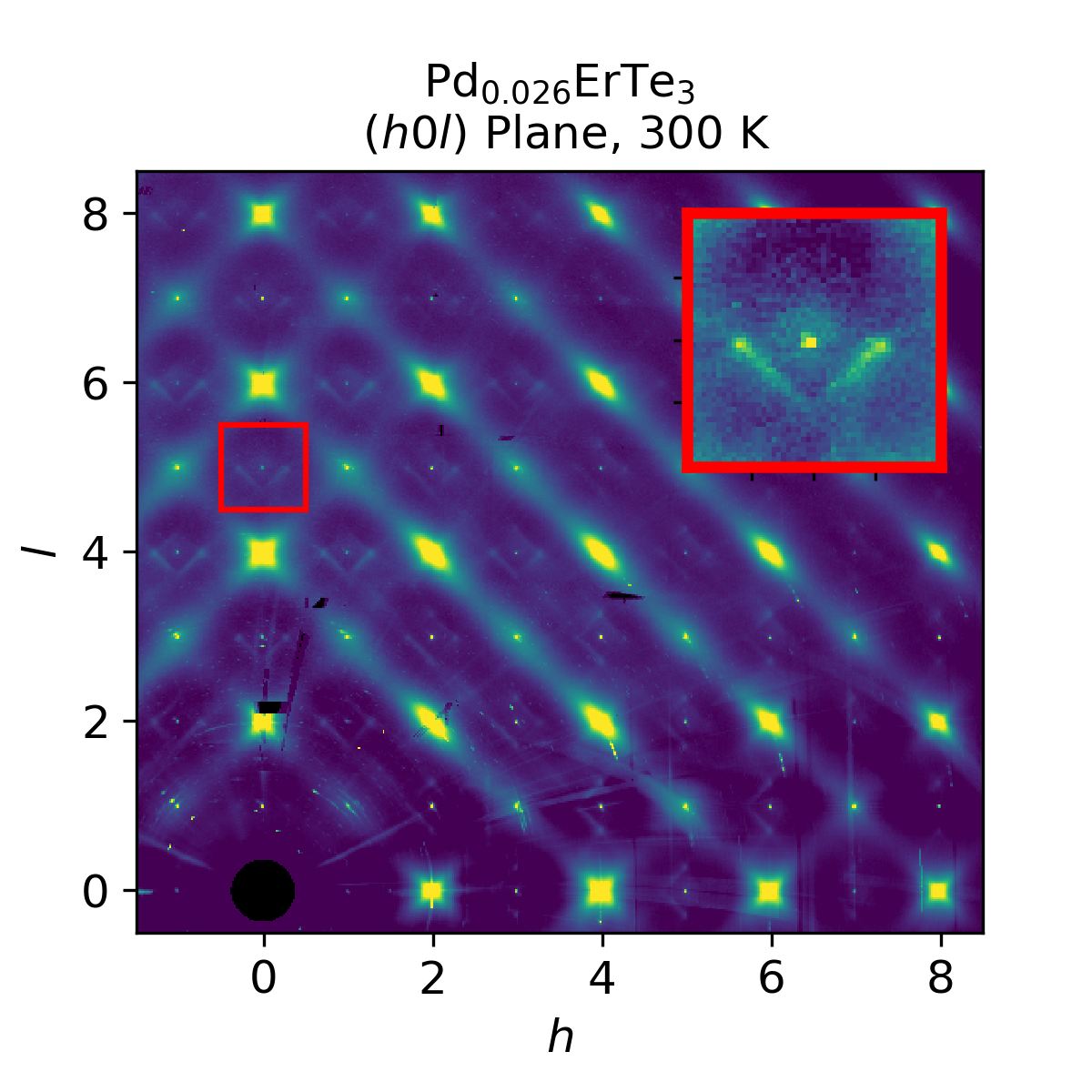}
\caption[XRD Study of the CDW State in Pd$_x$ErTe$_3$]{\label{fig_4_1} XRD image at room temperature of the $(h,0,l)$ plane of Pd$_{0.026}$ErTe$_3$. In addition to the Bragg peaks, significant diffuse scattering is evident. The large red box shows a zoomed-in version of this scattering in the small red box. Inspection of the scattering in other Brillouin zones reveals broad superlattice peaks along both the c-axis and a-axis directions, corresponding to short-range CDW correlations in both directions. These are connected by diagonal lines of diffuse scattering.} 
\end{figure}

The measurement also reveals what appears to be additional scattering peaks in the 2$\%$ sample. These peaks are in fact slices of broader diffuse scattering induced by palladium intercalation. This diffuse scattering can be viewed more completely in a $(h,l)$ scattering plane, as shown in Figure \ref{fig_4_1}. Remnants of the CDW superlattice peaks are still evident in the sample, despite the temperature being well above the critical temperature of the pristine material (268 K). Broad superlattice peaks are observed along both the c-axis and a-axis directions, connected by lines of diffuse scattering to form 'diamond' shapes. A similar pattern of diffuse scattering was previously observed in transmission electron microscope (TEM) studies of intercalated samples \cite{Straquadine2019}), and was attributed to the momentum dependence of the underlying electronic susceptibility (i.e. point-like defects can induce short range structural modulations, akin to Friedel oscillations, at wavevectors favored by the electronic susceptibility). Moreover, as shown in the inset, although superlattice peaks are observed along both the c-axis and a-axis directions, the diffuse scattering joining these points is clearly asymmetric around each reciprocal lattice point. 

%Since palladium intercalation does not alter the average structure of ErTe$_3$, a model of the pristine compound \cite{Singh2023} can be used as a starting point to predict the strain behavior of the disordered material. In the pristine compound, the materials behavior with strain will depend on the magnitude of the CDW order parameter and the value of the coupling between the CDW state and strain. XRD measurements of the parent and intercalated can be compared to evaluate how these terms change with disorder.

%Considering terms only to quartic order for the two CDW order parameters and neglecting gradient terms and the elastic energy, the contribution to the free energy density from the CDW is given by
%
%\begin{equation} \label{eq1}
%\begin{split}
%\Delta F & = r_a|\phi_a|^2 + u_a|\phi_a|^4 + r_c|\phi_c|^2 + u_c|\phi_c|^4 + g|\phi_a|^2|\phi_c|^2 \\
%& + \lambda_a^{xx}\epsilon_{xx}|\phi_a|^2+\lambda_a^{zz}\epsilon_{zz}|\phi_a|^2+ \lambda_c^{xx}\epsilon_{xx}|\phi_c|^2 + \lambda_c^{zz}\epsilon_{zz}|\phi_c|^2 
%\end{split}
%\end{equation}
%
%where $|\phi_a|$ and $|\phi_c|$ describe the amplitude of the two incommensurate CDW states and $\lambda_i^{jj}$ are strain coupling constants between the CDW state $|\phi_i|$ and the strain $\epsilon_{jj}$. To apply this model to the disordered sample, first it must be considered if any of the terms in the model, $|\phi_i|$ or $\lambda_i^{jj}$, change significantly with disorder. 

\begin{figure}
\centering
\includegraphics[scale=0.7]{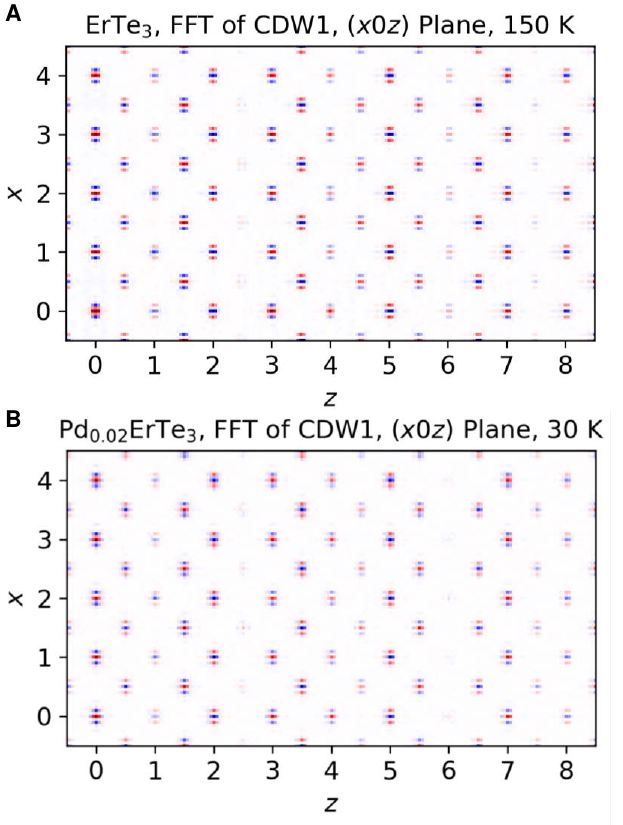}
\caption[Effect of Disorder on the CDW Order Parameter in Pd$_x$ErTe$_3$]{\label{fig_5_2} Displacement correlation maps of (A) parent and (B) intercalated samples at comparable temperatures relative to the observed CDW transition/ crossover. Values of $(x,z)$ are given in lattice units with each point representing a Te atom at that interatomic vector. The central peak at each interatomic vector describes the nominal position of the Te atom at that site, while the two side peaks give the value of the displacement of the Te atom in the CDW phase, roughly 0.05 lattice units for both samples. The color scale describes the correlation of the displacements, with red indicating a positive correlation and blue a negative correlation.  } 
\end{figure}

%The CDW order parameter term is evaluated first by observing directly the displacement of the Te atoms in the ordered phase. 
In Figure \ref{fig_5_2}, correlation maps of the displacement of Te atoms in the low-temperature phase for both a parent and intercalated sample are presented. To create these images,  the c-axis CDW peak was isolated from the volume of scattering data (used to produce Fig. \ref{fig_4_1}) by using Gaussian interpolation to estimate the background underneath reciprocal space positions $h = integer, l \pm 0.28$ and $h = integer, l \pm 0.40$. A Fourier transform was then applied to the resulting intensities to produce a three-dimensional pair distribution function resulting from the intensities at the CDW reciprocal lattice vectors. The displacement of the Te atoms was then determined by subtracting these positions from the average position of the atom in the high-temperature phase. To focus on the c-axis oriented CDW a Fourier filter was applied such that only the c-axis modulations were included. This produces a 3D-$\Delta$PDF \cite{Weber:2012en} of the isolated \textit{c}-axis CDW peaks. In the images shown in Figure \ref{fig_5_2}, each point in the $(x, z)$ plane represents a Te atom at an interatomic vector with that value of $(x, z)$. Each Te site is characterized by a three-peak structure; the central peak refers to the average position of the Te atom and the side peaks give the displacement of the Te atom in the ordered phase. The CDW is a transverse wave, hence a \textit{c-}axis wave is composed of displacements along the \textit{x-}direction. Red peaks indicate the displacement is positively correlated with that of the origin. Blue peaks indicate negative correlation, while the absence of coloring indicate no correlation between the displacements at the two sites. Comparing the two images, the magnitude of the displacement of the Te atoms is similar for the pristine and intercalated samples. In other words, although Pd intercalation introduces diffuse scattering, and necessarily changes how the CDW phase is correlated in real space, the atomic displacements that characterize the short-range CDW correlations look remarkably similar in the two cases. At least locally, far below $T_{CDW}$ the CDW order parameter, deduced here from local atomic displacements, in both samples (ErTe$_3$ and Pd$_{0.02}$ErTe$_3$) is very similar. 

\subsection{XRD as a Function of Anisotropic Strain}
\begin{figure}
\centering
\includegraphics[scale=0.65]{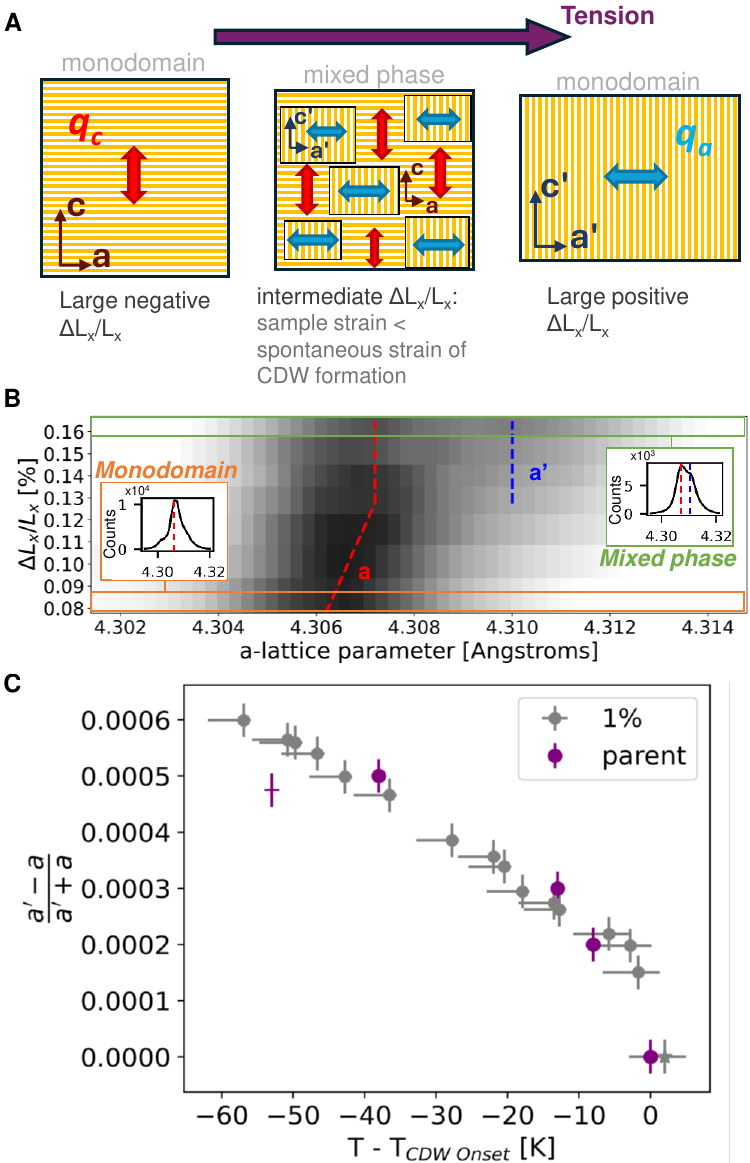}
\caption[Spontaneous Strain of CDW State in Pd$_x$ErTe$_3$]{\label{fig_4_3} (A) Schematic diagrams illustrating the effect of externally-induced strain on the CDW state of ErTe$_3$. A mixed phase region is anticipated and observed for intermediate values of the applied strain, as described in the main text. (B) Gray scale plot of the intensity of the (2,20,0) Bragg peak of a sample of 1$\%$ intercalated Pd$_x$ErTe$_3$ as a function of applied strain (vertical axis) and H (expressed here as the lattice parameter). Data were taken at 120K. Darker colors correspond to higher intensity. Dashed lines indicate the peak position as a function of strain. In the mixed phase, two peaks appear in the data due to the presence of domains. Insets show representative line cuts of the data in the monodomain (left inset) and mixed phase (right inset) regions, clearly showing a single peak vs double peak feature respectively. (C) Relative difference between the a-axis lattice parameters for the two types of CDW domains. Data shown in gray (purple) are for intercalated (pristine) samples respectively. Larger uncertainty in the orthorhombicity is observed for data collected with higher energy x-rays. Temperature error plotted for the intercalated sample is due to a different thermometer set-up resulting in a larger lag between recorded temperature and actual sample temperature. The temperature axis is offset relative to the temperature at which superlattice intensity is observed in each sample (T$_{CDW}$ onset)} 
\end{figure}

\begin{figure}
\centering
\includegraphics[scale=0.6]{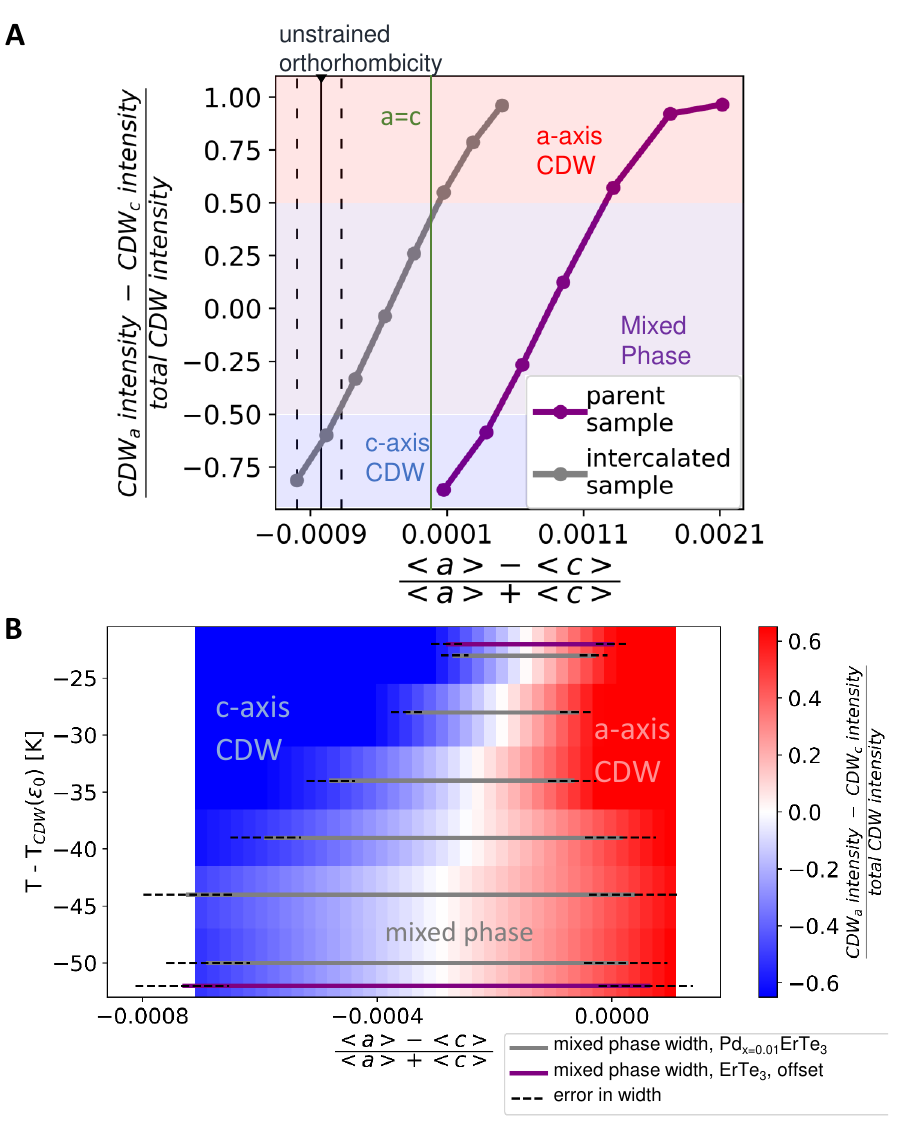}
\caption[Strain Switching of CDW State in Pd$_x$ErTe$_3$]{\label{fig_4_4} (A) Difference between integrated intensities of (1,27,q$_{CDW}$) peak and (q$_{CDW}$,27,1) peak, normalized by total CDW superlattice peak intensities at T- T$_{CDW}(\epsilon_0)$ = -50K, where $\epsilon_0$ refers to the unstrained sample, plotted versus induced orthorhombicity for both parent (purple data) and 1$\%$ intercalated (gray data) samples. (B) Relative intensity of c-axis and a-axis superlattice reflections (shown as a color scale) as a function of temperature and strain. The data delineate the boundaries of a phase diagram for the material, showing regions of c-axis CDW (in blue), a-axis CDW (in red), and mixed domain between these limits. The mixed domain phase is defined here as the region where neither CDW state exceeds 75$\%$ of the total CDW intensity. Gray lines indicate the mixed phase region for each measured temperature. For comparison, similar data are shown for the parent compound, as purple lines, determined in a similar manner. To facilitate comparison, these data were offset horizontally since, as demonstrated in (A), much larger strains are needed to reorient the primary CDW wavevector in the parent compound. The width of the mixed phase region is similar for both cases. } 
\end{figure}

%Having established that the magnitude of the atomic displacements associated with the CDW appears to be unaffected by intercalation, the next term to consider is the CDW state's coupling to the lattice or in other words the spontaneous strain induced by the CDW state. XRD study is combined with strain tuning to approach this question. 
Figures \ref{fig_4_3} and \ref{fig_4_4} summarize an x-ray study of a sample with nominally 1$\%$ palladium intercalation as a function of strain and temperature. The schematic in Figure \ref{fig_4_3}A illustrates the phenomenology that is generally observed as a function of strain in this material which is similar to that of pristine ErTe$_3$ \cite{Singh2023}. Specifically, as tension is applied along the \textit{a-}axis, the wavevector of the CDW is reoriented from along the c-axis direction to along the a-axis direction. A mixed domain phase is observed for intermediate strains, within which the sample can internally relax applied strains by adopting domains of the two CDW states. This is the purple region in the inset of Figure 1. In this mixed domain phase, as shown in Figure \ref{fig_4_3}B, structural Bragg peaks split since the lattice parameter along the \textit{a} or \textit{c} axis has different values in the two domains. The data shown in Figure \ref{fig_4_3}B were taken at a temperature of 120 K, well below the apparent onset of CDW order in a 1$\%$ intercalated sample.  

Strain causes the CDW wavevector to change direction. However, the crystal axes do not rotate along with the CDW (a consequence of the enforced orthorhombicity associated with the glide plane in the \textit{Cmcm} crystal structure). The lattice parameters change upon entering the CDW state (stretching along the direction of the CDW wavevector, while shrinking in the transverse direction [refs]). Hence, the mixed phase regime requires some care in labelling of lattice parameters. We adopt a convention in which unprimed symbols \textit{a} and \textit{c} refer to the lattice parameters in the c-axis CDW state (left schematic in Figure \ref{fig_4_3}), while primed symbols \textit{a'} and \textit{c'} refer to the same lattice parameters in the a-axis CDW state (right schematic in Figure \ref{fig_4_3}). In the monodomain regions, strain causes a uniform shift in these the lattice parameters. However, in the mixed phase region, peaks labelled by the same Miller indices are split. The example shown in Figure \ref{fig_4_3}B shows the (2,20,0) Bragg peak, with clear splitting in the mixed phase region. In this regime, changing the externally-induced macroscopic strain simply redistributes the population of the two CDW domain types, while the lattice parameters of each domain remain unchanged (Figure \ref{fig_4_3}B and ref \cite{Singh2023}). 

In the mixed phase region, the difference between \textit{a} and \textit{a'} provides information about the spontaneous strain induced by the CDW state. If the material were purely tetragonal, then the lattice distortion driven by the \textit{c-}axis and \textit{a-}axis CDW states would be the same along the two equivalent axes and hence given by exactly half of the difference between \textit{a} and \textit{a'}. In the orthorhombic case, however, since the two CDW states are distinct, no such symmetry relates the spontaneous strain induced by each CDW state. Nevertheless the difference between \textit{a} and \textit{a'} still provides a meaningful indication of the overall magnitude of the two spontaneous strains. This difference is plotted as a function of temperature for both the parent and intercalated sample in Figure \ref{fig_4_3}c relative to the onset temperature of the ordered state in the respective samples. As expected, this difference grows as the sample is cooled, consistent with a growing CDW order parameter. Of note, there is a striking similarity in the value of this difference for the parent and intercalated samples. This demonstrates that the spontaneous strain induced by the CDW states is remarkably similar in both samples, consistent with the earlier observation (Figure 5) that the local atomic displacements are also similar for the pristine and intercalated samples.

Next, we comment on the intensity of the CDW superlattice peaks themselves, as a function of strain. In Figure \ref{fig_4_4}A, the difference between the intensity of the CDW superlattice peak observed along the H direction and that observed along the L direction normalized by the total CDW intensity is plotted. As strain is applied, this parameter goes from negative to positive values as the direction of the CDW wavevector is rotated. The data in this figure is plotted as a function of the average sample orthorhombicity, \small $\dfrac{<a> -<c>}{<a>+<c>}$.\normalsize $\ $ Here the sample is characterized as being in the mixed domain phase if neither of the CDW states exceeds 75$\%$ of the total intensity of the CDW superlattice peaks. This type of measurement was repeated over a range of temperatures to create the phase diagram shown in Figure \ref{fig_4_4}B. Here, the mixed phase "dome" opening up below the CDW transition can be clearly observed (this is the purple region shown schematically in the inset of Figure 1). Across a wide range of temperatures, measured relative to the onset of CDW order, the width of the mixed phase region is similar for the parent and intercalated samples, further corroborating that the spontaneous strain induced by the CDW state is similar in both samples.

%These measurements in total strongly suggest that the spontaneous strain induced by the CDW state is similar in the disordered samples or in other words the coupling between strain and the CDW state, $\lambda_i^{jj}$, is unaffected by palladium intercalation. The similarity of the local CDW order parameter and the persistence of the orthorhombic distortion driven by the disordered CDW state could be understood by considering how intercalation disrupts the CDW order. By introducing both phase disorder and dislocations, intercalation disrupts the long range order of the CDW state. STM studies however reveal that robust, short range CDW correlations persist uniformly across the sample suggesting the presence of nematic, directional order on much longer length scales \cite{Fang2020}.

\begin{figure*}
\centering
\includegraphics[scale=0.725]{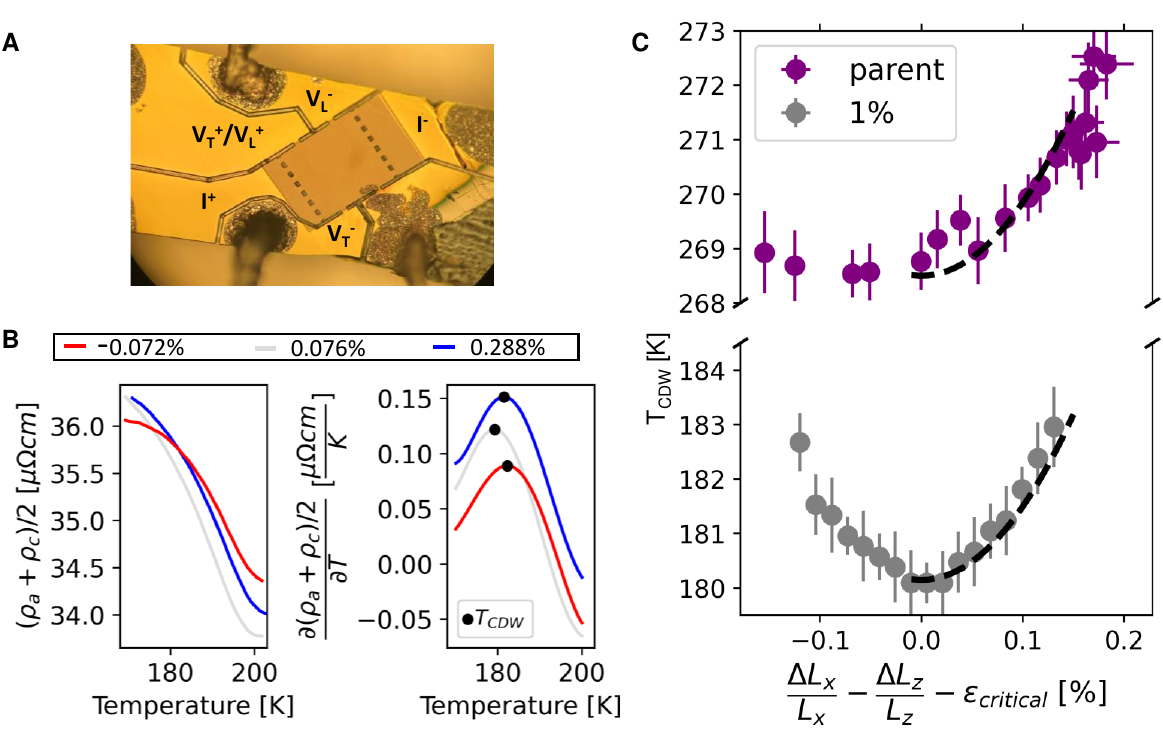}
\caption[Strain Tuning of T$_{CDW}$ in Pd$_x$ErTe$_3$]{\label{fig_4_5} (A) Image of microstructured device used to perform transport measurements of the 1\% Pd intercalated sample of Pd$_x$ErTe$_3$ under strain. The contact labeled $+/-$ V$_L$ correspond to the longitudinal voltage and give a direct measurement of $(\rho_a +\rho_c)/2$ while those labeled  $+/-$ V$_T$ correspond to the transverse voltage and give a direct measurement of $(\rho_a -\rho_c)/2$. In panel (B), the left panel plots the longitudinal resistivity versus temperature for three values of offset strain. The temperature derivative of this data is shown in the right panel, from which the onset temperature of the CDW is determined from the maximum of the derivative, indicated by the black dot for each curve. (C) T$_{CDW}$ as a function of applied strain for both the parent (purple data) and intercalated (gray data) samples, as determined from transport measurements. For each data set, the x-axis has been offset by the critical strain value such that the bicritical point in both datasets corresponds with 0$\%$ strain to facilitate comparison. Critical strain values, estimated based on the minimum of the these curves, were 0.17\% and 0.11\% for the pristine and intercalated compounds respectively. The black dashed curve plotted is the same in both datasets just vertically offset, demonstrating both samples have similar slopes for positive strains. } 
\end{figure*}

\subsection{Elastoresistivity Measurements}

The diffraction measurements presented thus far clearly demonstrate that the primary CDW wavevector can be reoriented with anisotropic strain in the intercalated sample, much like in the parent compound \cite{Singh2023}. Hence it is possible that there is a region in strain-temperature phase space where signatures associated with the vestiges of the two CDW phase transitions approach each other. In particular, even if the continuous phase transitions associated with the onset of CDW order are smeared due to disorder, the presence of a mixed phase regime clearly indicates the persistence of a first-order phase transition separating the disordered c-axis and disordered a-axis CDW states. This line of first-order phase transitions can still end in a critical point, which would play the same role as the putative bicritical point for the pristine compound. In this section we use a combination of resistivity and elastoresistance measurements to track the evolution of the phase diagram as intercalation is introduced. We use a microstructured transport device (a representative version of which is pictured in Figure \ref{fig_4_5}A) to perform high precision measurements of the resistive anisotropy while applying large strains to the sample. The sample shown has $x=1\%$ Pd intercalation.

We first explore the evolution of T$_{CDW}$ with strain for the disordered sample. As previously noted, in the presence of disorder this temperature cannot mark a true continuous phase transition to an incommensurate CDW state. Nevertheless, for small concentrations of Pd intercalation, the same general features of the CDW transition are still observed. To this point, even samples of "pure" ErTe$_3$ are not perfectly pristine. Crystal growth performed at finite temperatures necessarily results in some concentration of defects. Therefore, if considering only small concentrations of Pd, ErTe$_3$ and Pd$_x$ErTe$_3$ are likely fundamentally similar, only distinguished by the quantity of disorder. Hence, if it is possible to identify a CDW transition temperature in ErTe$_3$, it is reasonable to identify that same feature in Pd$_x$ErTe$_3$. Therefore, for simplicity, we label the features observed in resistivity measurements as T$_{CDW}$ for the disordered compound in this section. Measuring between the $V_L +/-$ contacts labeled in \ref{fig_4_5}A, gives values of the longitudinal resistivity, $(\rho_a +\rho_c)/2$, plotted in Figure \ref{fig_4_5}B for various values of applied strain. This data demonstrate a clear increase when passing through the phase transition, just as is observed for ErTe$_3$. Taking the derivative of this data, a peak can be identified as T$_{CDW}$. 

In figure \ref{fig_4_5}B, the temperature-dependence of the resistivity is presented for three representative strains, revealing that the value of T$_{CDW}$ is minimized at an intermediate value of applied strain presumably corresponding to the putative bicritical point. The full strain dependence of T$_{CDW}$ in the disordered sample is shown in Figure \ref{fig_4_5}C as well as that for a pristine sample. Each curve is horizontally offset such that its minimum lies at 0$\%$ strain. For positive antisymmetric strains (i.e. $\dfrac{\Delta L_x}{L_x} - \dfrac{\Delta L_z}{L_z} - \epsilon_{critical}>0\%$), the change in T$_{CDW}$ with strain for the intercalated sample is similar to that for the parent sample, again supporting that the CDW state's coupling to strain is unaffected by low levels of intercalation. However, in contrast to the pristine sample, the intercalated sample has much more symmetric behavior for positive and negative strains. The absence of a sharp ``V"-shaped minimum of T$_{CDW}$ as a function of strain is tentatively attributed to strain inhomogeneity which is averaged over in these transport measurements. 

Next, the in-plane resistive anisotropy is studied as a function of temperature. The sign of $(\rho_a - \rho_c)/2$ through the phase transition has been studied previously for unstrained samples, and understood based on the angular dependence of the Fermi velocity \cite{Sinchenko2014}. In particular, the resistivity grows larger along the direction that is transverse to the CDW wavevector. Thus, for the c-axis CDW state, $(\rho_a - \rho_c)/2$ is positive. Similarly, we have previously shown that for the pristine case of ErTe$_3$, the anisotropy changes sign when the CDW wavevector is rotated to lie along the a-axis \cite{Singh2023}. Thus the resistive anisotropy informs as to the balance of domains in the mixed phase regime, as well as the absolute value of the resistive anisotropy for mono-domain states. 

\begin{figure}
\centering
\includegraphics[scale=0.75]{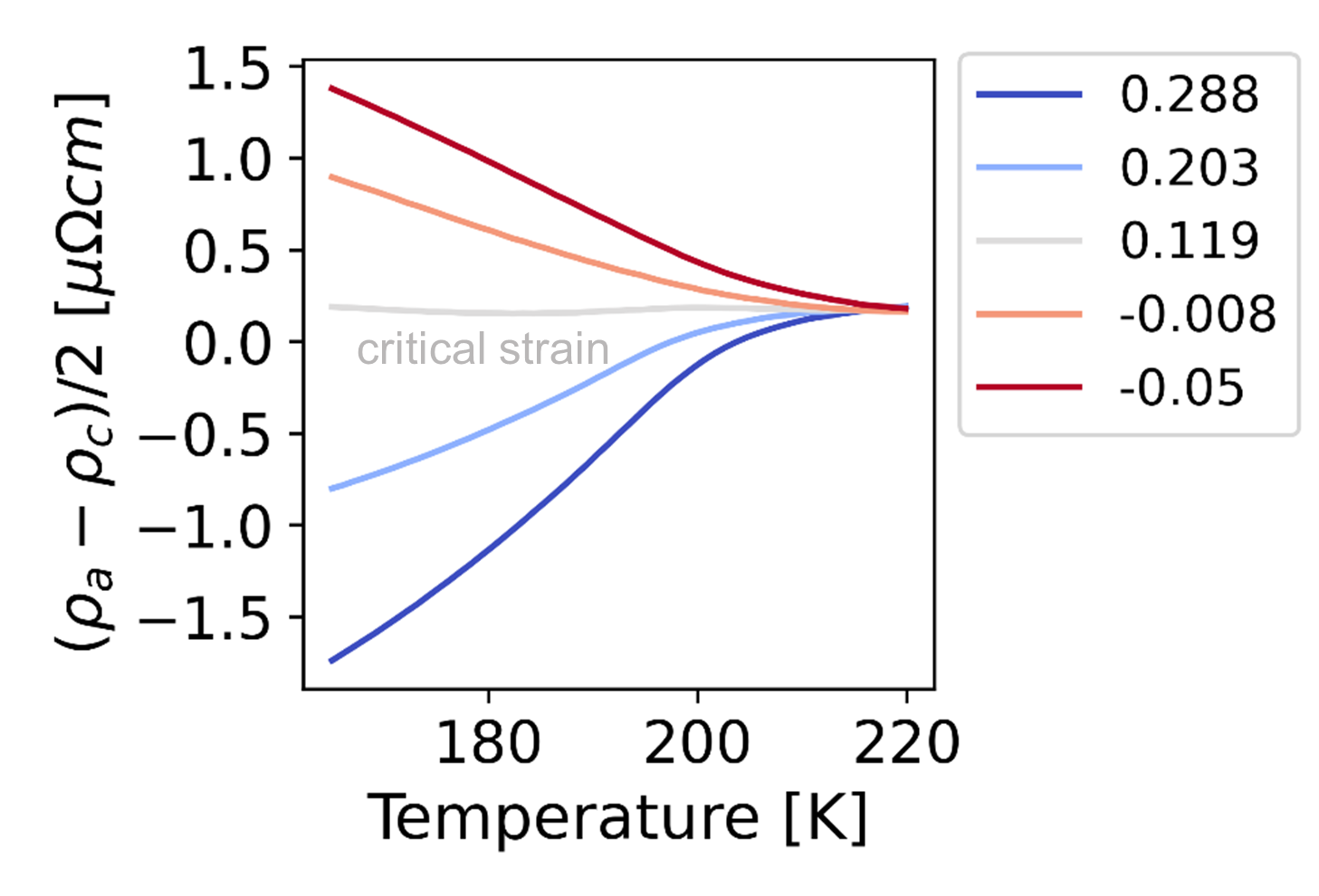}
\caption[Resistive Anisotropy in Pd$_x$ErTe$_3$]{\label{fig_4_6} The in-plane resistive anisotropy $(\rho_a - \rho_c)/2$ of of the 1\% Pd intercalated sample of Pd$_x$ErTe$_3$ as a function of temperature plotted for several representative values of applied strains through the critical temperature. Legend values indicate the average applied strain $\Delta L_x/L_x - \Delta L_z/L_z$ as a percentage for that temperature sweep. For compressive strains, domains with the c-axis CDW dominate and the resistive anisotropy is positive. For tensile strains, domains with the a-axis CDW dominate, and the resistive anisotropy is negative. The anisotropy does not saturate for the largest values of the applied strain, indicating either that these strains are not sufficient to fully detwin the material or that the material has a large elastoresistance even in the monodomain state.} 
\end{figure}

We measure the temperature-dependence of the transverse voltage ($V_T +/-$) in the device (Figure 8A) for different offset strains. This data is presented in Figure \ref{fig_4_6}. Overall, at high temperatures above the transition, the resistive anisotropy is small, but nonzero since the two in-plane axes are distinct. Below the CDW onset, the resistive anisotropy grows to either positive or negative values depending on which axis hosts the CDW state, which is controlled via strain. Similar behavior has already been seen for pristine ErTe$_3$ and TmTe$_3$ \cite{Singh2023, Straquadine2020}, though the magnitude of the resistive anisotropy for these intercalated samples is considerably smaller than for the pristine compounds. Notably, for the strain state corresponding to the minimum in T$_{CDW}$, the resistive anisotropy is unchanged through the CDW transition, indicating approximately equal domain populations for this value of strain. 

\begin{figure}
\centering
\includegraphics[scale=0.5]{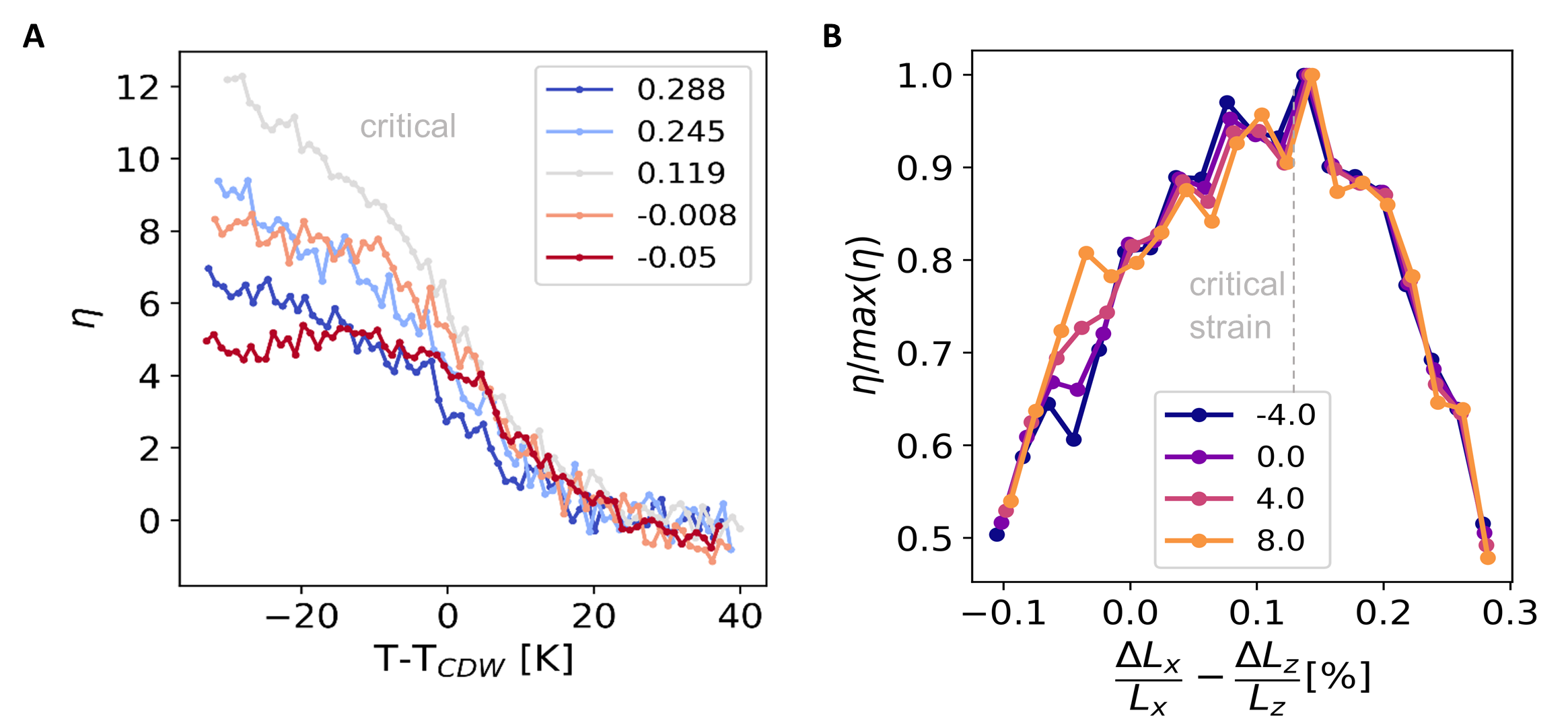}
\caption[Nematic Elastoresistance in Pd$_x$ErTe$_3$]{\label{fig_4_7} (A) The nematic elastoresistance $\eta$ of the 1\% Pd intercalated sample of Pd$_x$ErTe$_3$ as a function of relative temperature, plotted for several representative values of applied strain. Similar to the data shown in Figure \ref{fig_4_6}, the reported values of strain are average values of $\Delta L_x/L_x - \Delta L_z/L_z$  over that temperature sweep. (B) The nematic elastoresistance as a function of applied strain for fixed values of distance to criticality, T-T$_{CDW}$ (values listed in legend). Data are normalized by the peak value at each temperature to facilitate comparison.
}
\end{figure}

The nematic elastoresistance $\eta$ of the intercalated sample, defined as:
\begin{align}
    \eta \equiv \frac 1 {(\rho_a+\rho_c)} \ \frac {\partial (\rho_a-\rho_c)}{\partial (\epsilon_{xx}-\epsilon_{zz})}
    \label{eta}
\end{align}
is presented in Figure \ref{fig_4_7}. In Figure, \ref{fig_4_7}A, $\eta$ is plotted as a function of relative temperature, T-T$_{CDW}$, for several values of offset strain. In each measurement, $\eta$ begins to increase well above T$_{CDW}$ and grows increasingly faster approaching T$_{CDW}$. Additionally, there is an intermediate value of applied strain for which the value of $\eta$ grows the largest. This can be seen more clearly by plotting $\eta$ as a function of applied strain for different values of relative temperature as is done in Figure \ref{fig_4_7}B. A clear maximum in the nematic elastoresistance $\eta$ is observed corresponding to the strain value previously identified as the putative bicritical point. Though the exact shape of this peak is dependent on the definition of T$_{CDW}$ as a function of strain, the existence of a clear peak is indifferent to this definition.

\begin{figure}
\centering
\includegraphics[scale=0.6]{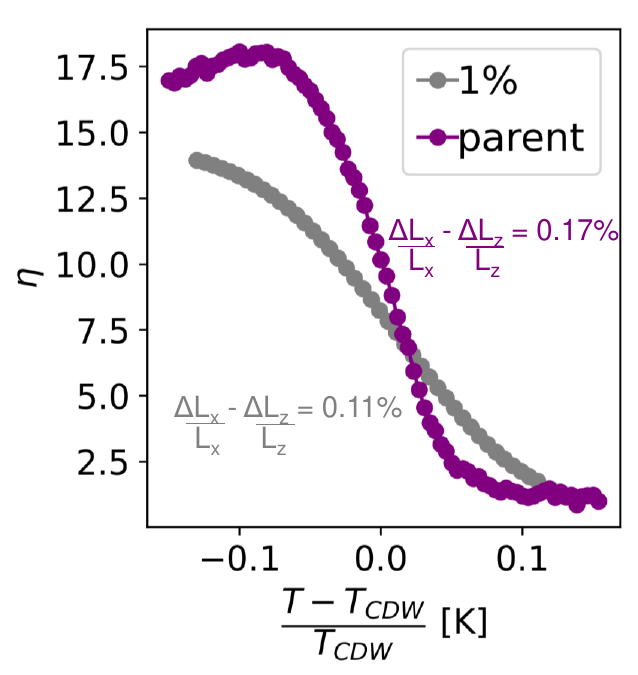}
\caption[Comparison of Nematic Elastoresistance to ErTe$_3$]{\label{fig_4_8} The nematic elastoresistance plotted as a function of the reduced temperature for the intercalated and parent sample at each samples’ observed critical strain. The value of the critical strain for each sample is indicated on the plot.} 
\end{figure}

In Figure \ref{fig_4_8}, $\eta$ at the critical strain is plotted for both the parent and disordered samples as a function of reduced temperature, $\dfrac{T-T_{CDW}}{T_{CDW}}$ to facilitate comparison. Close to the critical point, the elastoresistive response of the intercalated sample is suppressed relative to the pristine compound. In both cases, the nematic elastoresistivity rapidly drops upon increasing temperature. However, the suppression is less rapid for the intercalated sample, such that the enhanced elastoresistivity associated with the bicritical point stretches over a wider range of reduced temperatures.  

We also note that the value of the critical strain extracted from these elastoresistivity measurements, noted in grey and purple text in Figure \ref{fig_4_8} for the intercalated and pristine samples respectively, is considerably smaller for the intercalated sample. This is consistent with the XRD measurements described earlier (Figure \ref{fig_4_4}A), which provided a more direct measure of the strain. Additionally, as also determined by XRD, for neither the parent nor intercalated sample is this critical strain dictated by the condition of the the $a$ and $c$ lattice parameters being equivalent.

\begin{figure}
\centering
\includegraphics[scale=0.575]{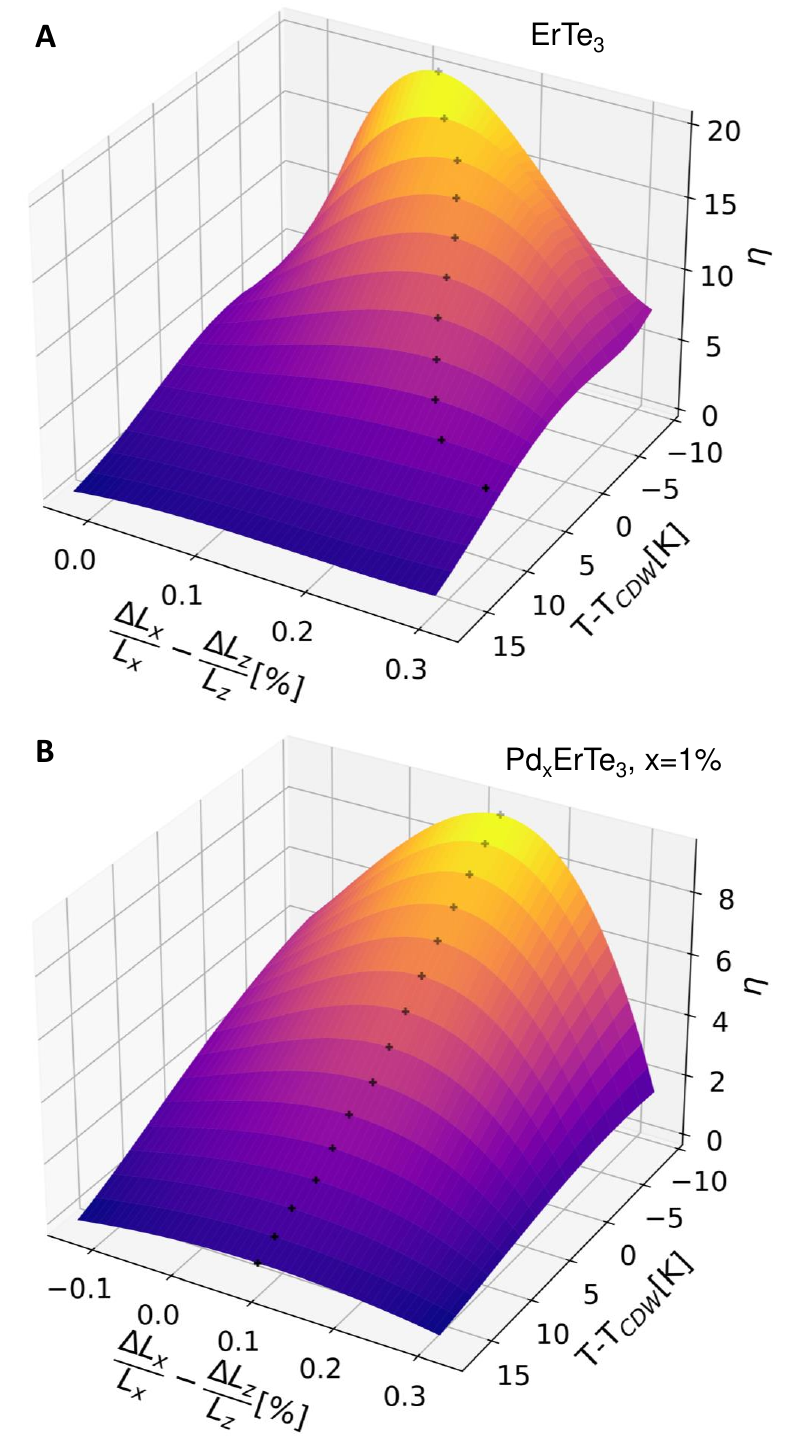}
\caption[Temperature- Strain Phase Diagram of Nematic Elastoresistance]{\label{fig_4_9} The nematic elastoresistance $\eta$ plotted as a function of relative temperature and applied strain for (A) the parent and (B) intercalated sample respectively. Black markers indicate the position of the maximum value in $\eta$ to illustrate its evolution with temperature. The data are much more symmetric for the intercalated sample relative to the pristine case.} 
\end{figure}

In Figure \ref{fig_4_9} the nematic elastoresistance for both the parent and disordered samples are plotted over the complete temperature-strain phase space. The suppression and broadening of the nematic elastoresistance in the intercalated sample relative to the pristine sample is clearly visible. Additionally, this plot allows comparison of the temperature evolution in the behavior of the critical strain, identified by the maximum in $\eta$. In the intercalated sample, the value of the critical strain does not change considerably with temperature and the nematic elastoresistance is significantly more symmetric around this point. In contrast, for the parent sample, the data is less symmetric about the critical strain and further from the bicritical point, the value of the critical strain changes considerably. 

\begin{figure}
\centering
\includegraphics[scale=0.815]{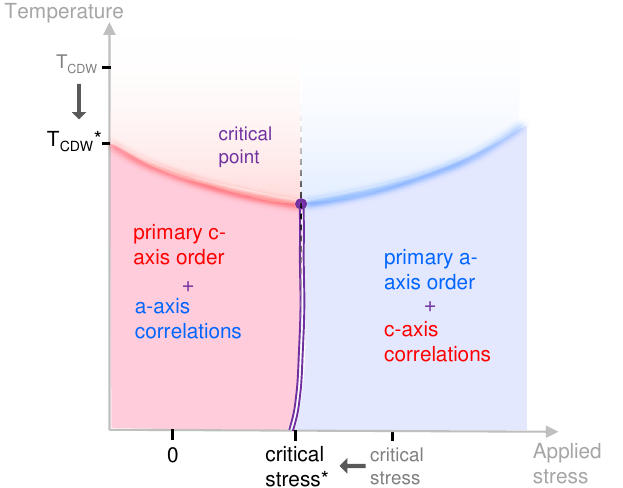}
\caption[Proposed Phase Diagram for Disordered Sample]{\label{fig_13} Proposed phase diagram for Pd$_x$ErTe$_3$ as a function of stress. The continuous phase transition associated with the formation of the CDW state is rounded to become a cross-over, though the characteristic signatures associated with the onset of long-range CDW order are still evident. These features (marked by T$_{CDW}^*$) are suppressed in temperature relative to the pristine compound. The present work is agnostic as to whether this corresponds to a Bragg glass transition. Signatures of short-range CDW correlations are present well above the suppressed cross-over temperature (light shading). A first-order phase transition (purple double line) still separates regions of the phase diagram in which the primary CDW wavevector is oriented along the c-axis (red region) and the a-axis (blue region), but short-range correlations exist in both directions on both sides of the phase transition. This line of first order phase transitions must end in a critical end point. Dashed line eminating from this end point indicates value of stress that maximizes the nematic elastoresistance, and is almost vertical. Comparing the phase diagrams of this disordered system with that of the pristine compound (Figure 1), the emergent behavior proximate to the critical endpoint is apparently more symmetric farther away from this point than is found for the bicritical point in the pristine compound, despite the fact that the material is still fundamentally orthorhombic.} 
\end{figure}

\section{Discussion}
We start by briefly reiterating the key observations associated with Pd intercalation in ErTe$_3$ that have been revealed by the present study. These are:\\
(1) The average crystal structure of Pd$_x$ErTe$_3$ belongs to the same orthorhombic space group as that of the pristine compound, and the in-plane lattice parameters have a similar (small) anisotropy; \\
(2) At low temperatures, in the CDW state, the CDW-induced spontaneous strain is similar for the Pd intercalated sample as for the pristine sample. Similarly, the atomic displacements are similar;\\
(3) Diffuse scattering at high temperatures reveals short range CDW correlations in both in-plane directions, presumably pinned by the Pd intercalants (which we correspondingly loosely refer to as the source of disorder); \\
(4) Signatures of a CDW phase transition are observed in both resistivity and x-ray diffraction. By tracing these as a function of Pd concentration these can be smoothly connected to those which are found for the pristine ErTe$_3$. From a theoretical perspective, true long-range order associated with an incommensurate CDW is precluded in the presence of disorder \cite{Larkin1970,ImryMa1975}. These signatures therefore presumably correspond to either a cross-over or possibly to a phase transition to a related low $T$ phase - either a Bragg glass or a phase with vestigial discrete order of some sort; \\
(5) The characteristic temperature associated with this cross-over/transition is suppressed relative to the pristine compound; \\
(6) For samples cooled below this characteristic temperature, varying the strain reveals a first order phase transition, with a mixed phase region for intermediate strains. The critical strain associated with the end point of this line of first order phase transitions occurs for a smaller value of the strain than for the pristine compound; \\
(7) The phase diagram is more symmetric about this critical strain than it is for the pristine compound. Specifically, the critical/crossover temperature of the two CDW states is more symmetric, the nematic elastoresistance above the critical point is more symmetric, and the resistive anisotropy below the critical point is more symmetric. In other words, despite still being orthorhombic from a structural perspective, the material is electronically much more isotropic than is 
the pristine compound;\\
(8) And finally, the nematic elastoresistance is peaked upon approaching the critical end point, but it diverges more slowly and it extends over a wider range of temperature than in the pristine compound.

 Most of these observations can be heuristically/qualitatively understood as a consequence of the disorder that Pd intercalation introduces. 
 
First we consider the short range CDW correlations. As has previously been noted, point-like defects induce short-range correlations that reflect the underlying electronic susceptibility \cite{Straquadine2019}. The near-four-fold symmetry of the susceptibility \cite{JohannesMazin2008} necessarily means that the induced static short range correlations are also nearly four-fold-symmetric, as previously observed via electron diffraction \cite{Straquadine2019}, and here via diffuse x-ray scattering. 

The suppression of the onset of CDW order with increasing disorder is anticipated in the case that this corresponds to a vestigial order of some sort; the broken discrete symmetry can survive in the presence of disorder, but only up to some critical value \cite{Nie2014}. Similarly, if this corresponds to a Bragg glass transition, it is anticipated that beyond some critical disorder concentration the quasi-long-range order is destroyed, implying that the critical temperature is also suppressed with increasing disorder \cite{Fang2019}. Thus, both types of phase transition would naturally be suppressed as a function of disorder. In contrast, there is no similarly universal prediction for the disorder-dependence in the case that this corresponds to a cross-over, though the observed behavior is certainly not precluded. 

The apparent decrease in electronic anisotropy with increasing Pd intercalation, affecting the phase diagram, the resistivity anisotropy below the critical temperature and the elastoresistivity above the critical point, is consistent with expectations based on the averaging effects of scattering. A quantitative analysis would require more information than can be gleaned from the measurements presented here, but at least all of these physical quantities can, in broad terms, be anticipated to become more isotropic as the strength of disorder is increased.

Similarly, the progressive suppression as a function of increasing disorder of the divergence of the elastoresistivity approaching the critical point is consistent with expectations based on analysis of the random field Ising model \cite{Kuo2016}. Similar sub-Curie behavior has been seen in elastoresistivity measurements of chemically-substituted Fe-based superconductors \cite{Kuo2016}. 

Finally, we consider one of the most fascinating aspects of ErTe$_3$, the emergent tetragonality close to the bicritical point \cite{Singh2023}. This behavior is formally related to an emergent Z$_2$ symmetry associated with the meeting of two U(1) symmetric order parameters associated with the incommensurate CDWs \cite{Pandey2023}. For the disordered case of Pd$_x$ErTe$_3$, at a formal level there is no U(1) symmetry-breaking phase transition (because disorder must, at some level, pin the phase of the two CDWs \cite{Larkin1970,ImryMa1975}). However, it is intriguing to note that the phenomenology exhibited by Pd$_x$ErTe$_3$ is as if the intercalated sample exhibits an effectively tetragonal behavior for a larger range of temperature-strain phase space around the critical point than for the pristine ErTe$_3$. The extent to which the near-tetragonal behavior in this part of the phase diagram derives from critical fluctuations at the critical endpoint (for example, similar to at the endpoint of a line of nematic phase transitions \cite{Pandey2023}) relative to simply reflecting a more averaged electronic response due to disorder, is not immediately apparent, though both effects inexorably lead to a more isotropic response close to this special point in the phase diagram.

\section{Acknowledgments}
This work was supported by the Department of Energy, Office of Basic Energy Sciences, under contract DE-AC02-76SF00515. This research used resources of the Advanced Photon Source, a U.S. Department of Energy (DOE) Office of Science User Facility operated for the DOE Office of Science by Argonne National Laboratory under Contract No. DE-AC02-06CH11357. XMaS is a UK national research facility supported by EPSRC. We are grateful to all the beamline team staff for their support. A.G.S. was additionally supported by the National Science Foundation Graduate Research Fellowship Program under grant no. DGE-1656518 and the Department of Energy Office of Science Graduate Student Research (SCGSR) Program. M.D.B. was additionally supported by the Geballe Laboratory for Advanced Materials Postdoctoral Fellowship.

\bibliographystyle{unsrt_abbrv_custom}
\bibliography{apssamp.bib}

\appendix
\section{Fourier analysis of CDW peaks}

\begin{figure*}
\includegraphics{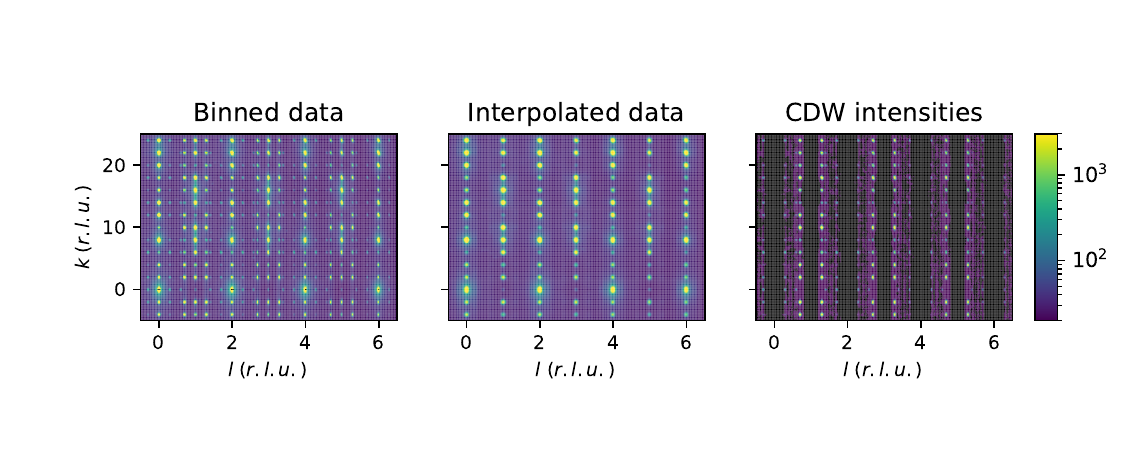}
\caption[Data Processing for FFT]{\label{punch_fill_fig} {To isolate the $c$-axis} CDW peaks, a punch-and-fill type algorithm was used over the CDW positions to estimate the underlying background. From the original data (left), $c$-axis CDW positions were removed and interpolated via Gaussian convolution, providing a background estimate at those positions (center). This was then subtracted from the original data, with all non-CDW positions set to zero (right). This volume was the input for the FFT.} 
\end{figure*}

The displacement correlations displayed in Figure \ref{fig_5_2} were derived from large but relatively low-resolution three-dimensional reciprocal space intensity maps using a Patterson-like method, where a Fourier transform is applied to the relevant intensities to generate a three-dimensional pair distribution function. The intensities chosen here were the expected first-order $c$-axis CDW peaks at $h = integer, l \pm 0.28$ and second-order $c$-axis CDW peaks at $h = integer, l \pm 0.40$, within a radius of $\approx0.15 $ \AA\ in the $hl$ plane. All $k$ values were used so that this method could also be applied to the diffuse scattering seen at these locations above $T_C$. To avoid the effects of background, data within these rods were first removed from the data and interpolated from the rest of the dataset, similar to the "punch-and-fill" procedure outlined in \cite{Krogstad:2019tc}. Here, the interpolated values are then subtracted from the original dataset, and all values outside these rods were set to zero, thus isolating the $c$-axis CDW intensities for Fourier analysis. Representative planes from each step are shown in Figure \ref{punch_fill_fig}. As in \cite{Krogstad:2019tc}, an outer product Tukey windowing function was applied prior to the Fourier transform to suppress spurious signals.

\end{document}